\documentclass[11pt]{article}
\usepackage[margin=1in]{geometry}
\usepackage[T1]{fontenc}
\usepackage[utf8]{inputenc}
\usepackage{lmodern,microtype}
\usepackage{amsmath,amssymb,booktabs,tabularx,array}
\usepackage[authoryear,round]{natbib}
\usepackage{graphicx,caption,enumitem}
\usepackage[colorlinks=true,allcolors=blue]{hyperref}
\usepackage{xurl}
\setlist{nosep,leftmargin=*}
\newcommand{\CI}[2]{[#1,\,#2]}
\title{KITE: Scaling Jev Population Experiments\\with Sparse Flagship Calibration}
\author{Hengyu Li\\The University of Tokyo\\\texttt{li-hengyu@g.ecc.u-tokyo.ac.jp}}
\date{September 23, 2026}
\hypersetup{pdftitle={KITE: Scaling Jev Population Experiments with Sparse Flagship Calibration},pdfauthor={Hengyu Li},pdfsubject={Human-referenced population simulation and sparse flagship calibration}}
\begin{document}
\maketitle
\begin{abstract}
KITE queries a typed behavioral kernel once per unique state, then executes populations of any size from the table with event-keyed randomness and common random numbers. An expensive flagship model is reserved for sparse paired anchors that estimate intervention effects. Measured human--model discrepancy is propagated as shared error into every conclusion. Population-experiment cost thus scales with unique states and anchors, while uncertainty is governed by evidence about people rather than Monte Carlo noise. On Epstein experiments with 9,070 participants, anchors covering 1.7\% of states reduced effect error by 41\% (absolute MAE reduction 0.0125). On 37 held-out SocSci210 experiments, 0.5--1.5\% anchor coverage raised captured decision gain from 0.27 to 0.39. The kernel passed content-fidelity criteria in all 15 new countries of a 16-country study. Shared discrepancy yielded retrospective coverage of 93\% and 96\% at nominal 80\% and 90\%, versus 29\% and 36\% from human sampling uncertainty alone. A million agents executed 20 tabulated steps in 0.9 seconds on a laptop. This architecture offers a route to screening candidate interventions before human trials, multi-country content audits, and uncertainty-aware policy comparison at the cost of a few thousand kernel calls with sparse flagship anchors. Property-specific evidence records connect each use to its validation scope, correction provenance, and uncertainty, making these applications auditable.
\end{abstract}

\section{Introduction}
A population experiment requires a model of how responses change when a situation changes. It also requires a way to distinguish computational precision from agreement with people. Cheap inference helps with the former; it cannot establish the latter. KITE (Kernel, Intervention anchors, Tabulated execution, Error propagation) addresses this distinction by combining a typed behavioral kernel, sparse flagship-model corrections, tabulated population execution, and an empirical model of human--model discrepancy.

The kernel is TypeSafe's Jev, pinned to \texttt{jev-1.13.0}, released on September 15, 2026. TypeSafe calls this a ``System One'' model: it maps text or structured state to predefined decision types rather than generating free-form language \citep{typesafe2026}. Here this product designation implies no decomposition of human cognition. We use it to predict a distribution over a described respondent's answers. Such an interface permits repeated, controlled population experiments in ordinary code, with separately specified states, treatments, outcomes, and random numbers.

Sparse calibration improves effect estimates and intervention choices while preserving inexpensive population execution. Our contributions are:
\begin{enumerate}
\item \textbf{Sparse flagship calibration with human-referenced evaluation.} In Epstein, effect MAE falls from 0.0305 to 0.0180 ($-41\%$; paired error reduction 0.0125, 95\% CI \CI{0.0064}{0.0167}). In SocSci210, 0.5--1.5\% anchor coverage raises captured gain from 0.268 to 0.386 (gain 0.119, \CI{0.014}{0.224} at three anchors).
\item \textbf{A measured allocation of behavioral capabilities.} Content-fidelity gates pass in all three test parts of a 16-country study, including all 15 new countries in both item halves. This supports using the cheap kernel for content response and sparse corrections for demonstrated intervention families.
\item \textbf{Shared discrepancy propagation.} On 37 unseen studies, retrospective coverage is 0.93/0.96 at nominal 0.80/0.90, compared with 0.29/0.36 for sampling-only intervals. Discrepancy is shared across agents rather than averaged away by population size.
\item \textbf{An implemented prediction--execution separation.} Fresh kernel predictions cost \$0.0227 per thousand in the scale run; tabulated execution of $10^6$ agents through 20 steps takes 0.90\,s. Evidence records attach validation scope and uncertainty to each output. A separately evaluated memory tier passes coherence and marginal-preservation criteria in six studies.
\end{enumerate}
The contribution is the implemented, evaluated combination. Frozen criteria distinguish primary, secondary, retrospective, and developmental evidence.

\section{Related work}
\paragraph{Agent simulators.}
Generative Agents combines memory, reflection, and planning to produce interactive social behavior \citep{park2023}; interview-grounded simulations extend this approach to representations of real respondents \citep{park2024}. Concordia supports agents acting in social, physical, and digital environments \citep{vezhnevets2023}. OASIS and AgentSociety develop infrastructure for larger social simulations \citep{yang2024,piao2025}. Their execution scale does not by itself establish intervention validity.

APS is a particularly close architectural comparison \citep{zheng2026}. The project specification describes adaptive prototypes, shadow audits, tail routing, and residual correction, with 10 million agents and approximately 381-fold fewer LLM calls, assessed against a full-LLM reference. We treat these as reported approximation results rather than evidence of agreement with human populations. KITE evaluates a complementary question: whether a computational approximation preserves decisions supported by human experiments and how residual behavioral error changes population conclusions. A good approximation can preserve a reference model's mistakes.

\paragraph{Synthetic respondents and behavioral prediction.}
Silicon sampling conditions language models on respondent descriptions \citep{argyle2023}. OpinionQA and GlobalOpinionQA examine whose opinions models represent \citep{santurkar2023,durmus2023}. Critiques show why agreement in aggregate answers does not establish valid variation, statistical relationships, or robustness to survey presentation \citep{bisbee2024,dominguez2024}. SocSci210 evaluates response distributions across social-science experiments \citep{kolluri2025}; \citet{ashokkumar2026} study prediction of experimental results. KITE distinguishes marginal fit, condition sensitivity, intervention choice, and within-person dependence. Its demographic-persona findings do not generalize to the richer interview representations of \citet{park2024}.

\paragraph{Simulation and uncertainty.}
Combining inexpensive and expensive models has a substantial multifidelity literature \citep{kennedy2000,peherstorfer2018}. Common random numbers reduce variance in contrasts, while conditional expectations can remove response-draw noise \citep{glasserman2004}. Calibration against observations requires a model--reality discrepancy term \citep{kennedy2001,brynjarsdottir2014}. We adapt these ideas to behavioral effect estimates; the flagship is an additional fallible model. The ODD tradition requires explicit states, scheduling, assumptions, and evaluation of fitness for purpose \citep{grimm2006,grimm2020}. Evidence records apply that discipline to individual reported properties rather than compressing them into a scalar trust score.

\section{The KITE architecture}
\subsection{Kernel contract and population}
For state $s=(d,h,z,a)$, containing a rendered persona $d$, content $h$, situation $z$, and condition $a$, the kernel returns $p_K(y\mid s)$ on a finite, explicitly labeled answer space. All arithmetic, control flow, weighting, and execution remain in code. Jev supports Choice, Score, and Noul primitives. The frozen \texttt{p3} contract asks one Noul probability per option, bundles these questions in one request, and renormalizes their outputs. The question is third-person and predictive: what answer would the described respondent give? It is not a request for the model's own opinion or for the normatively correct answer.

Question wording, option text, fixed field order, and model version belong to the contract and cache key. Numeric quantities are rendered as verbal categories where appropriate; option text replaces arbitrary letter labels, and phrasing-specific thresholds are not transferred between primitives. The task preserves condition-specific scales and labels, gives information in questionnaire order, and excludes outcome variables from personas. Absolute option judgments were selected on development studies: p3 expected calibration error was 0.027, compared with approximately 0.23--0.25 for relative-choice formulations. This empirical result does not imply that renormalization guarantees human calibration.

Personas are rendered from eligible microdata covariates, with population weights kept separately. Identical rendered states share a cache entry, even when they originate from different respondents. Demographic descriptions are population inputs, not validated identities. Externally supplied veracity labels define misinformation correction cells; the system does not infer ground truth merely by simulating a population.

\subsection{Prediction and execution}
The tabulated tier calls the kernel once per unique state and executes responses by inverse-CDF sampling. An event-keyed uniform variate is determined by the simulation seed, agent, step, and event; treatment branches use common random numbers without depending on call order. For a fixed table, a weighted expectation
\begin{equation}
\widehat\mu_a=\frac{\sum_i w_i\sum_y y\,p_K(y\mid d_i,h_i,z_i,a)}{\sum_i w_i}
\end{equation}
removes response-draw noise entirely. This Rao--Blackwell calculation still depends on the supplied population, table, and behavioral model.

The measured need for separation is concrete: 29.2\% of the nominally fresh states in the scale workload repeated an existing rendered state. The implementation provides a vectorized event-keyed engine and a scalar engine, alongside caching, throttling, budgets, and ledgers. Its measured large-population workload has fixed content exposure and no evolving network.

\begin{figure}[tp]
\centering\includegraphics[width=\linewidth]{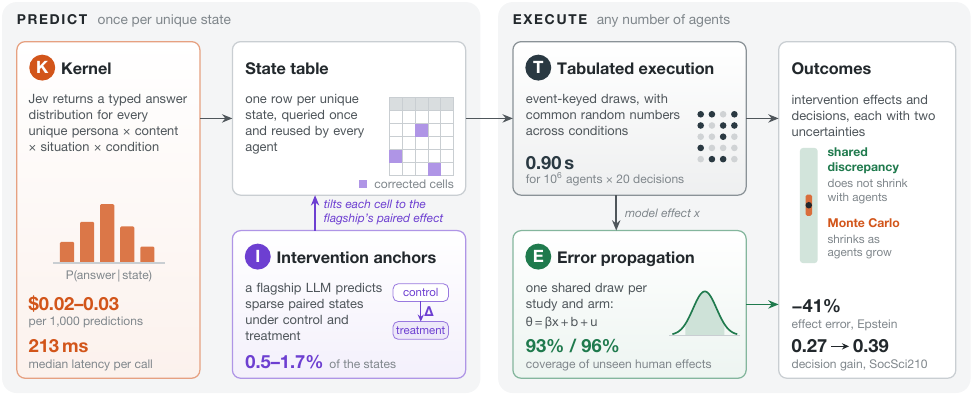}
\caption{KITE separates prediction from execution. PREDICT operates once per unique state; EXECUTE serves any number of agents. Colored cards bear letter badges: K, Kernel; I, Intervention anchors; T, Tabulated execution; E, Error propagation. Arrows connect the typed Jev kernel, white state table, execution, and white outcomes card. Paired control--treatment flagship anchors tilt highlighted table cells to the flagship's paired effect; each row is queried once and reused by every agent. Execution uses event-keyed draws with common random numbers across conditions, passing model effect $x$ through error propagation, $\theta=\beta x+b+u$, to outcomes. Outcomes distinguish shared discrepancy drawn once per study and arm from Monte Carlo error, which alone shrinks with more agents.}
\label{fig:architecture}
\end{figure}

\subsection{Sparse correction and its provenance}
For correction cell $g$, treatment $a$, and reference condition $0$, a flagship estimates the mean paired effect over anchor states $A_g$:
\begin{equation}
\widehat\delta^F_{ga}=\frac{1}{|A_g|}\sum_{i\in A_g}\bigl[m_F(s_i,a)-m_F(s_i,0)\bigr],
\qquad m^*_{ga}=\overline m_{K,g0}+\widehat\delta^F_{ga}.
\label{eq:target}
\end{equation}
Thus the flagship effect replaces the kernel effect. Adding the correction to the kernel's treated mean would count both effects. In Epstein, cells are veracity crossed with tertiles of control-arm $P(\mathrm{share})$, computed per wave and veracity at the persona--headline level. There are 12 anchor states per cell, evaluated under each relevant condition in separate batches.

The operator \texttt{kite.operators.tilt} transforms each kernel \emph{treated} distribution:
\begin{equation}
p^*_{iay}=\frac{\widetilde p_{iay}\exp(\alpha_{ga}y)}{\sum_{y'}\widetilde p_{iay'}\exp(\alpha_{ga}y')},
\qquad \frac{\sum_{i\in g}w_i\sum_y yp^*_{iay}}{\sum_{i\in g}w_i}=m^*_{ga}.
\label{eq:tilt}
\end{equation}
Here $\widetilde p$ is the normalized distribution after applying the frozen probability floor 0.005; $|\alpha|\leq10$. A one-dimensional solve matches the cell mean. If the target is unreachable or fewer than six paired anchors are available, the cell retains the uncorrected kernel predictions and reports its affected weight. A shared tilt matches a mean; it does not recover individual treatment effects or a full human response distribution.

The cell rule came from D0, an in-sample Arechar prototype. A global country-level tilt moved the prompt effect from $-0.060$ to between $-0.091$ and $-0.071$ across anchor budgets, worsening discernment. Veracity cells recovered the flagship's direction; finer baseline cells improved approximation without necessarily improving human fit. This is developmental evidence for representation choice, not another independent hybrid test. In SocSci210, D3 implements Equation~\ref{eq:target} directly at cell-mean level, using study--task blocks rather than veracity cells; it does not validate a distributional tilt on every survey scale.

Each operator records whether its source is a human estimate, a sparse flagship estimate, or a user-supplied sensitivity range. The measured corrections here are flagship-sourced. The broader provenance interface must not relabel them as experimentally estimated human effects.

\subsection{Discrepancy, behavioral tiers, and implementation status}
For study $s$ and nonreference arm $a$, let $x_{sa}$ be a model effect and $\theta_{sa}$ its human counterpart. The fitted discrepancy model is
\begin{equation}
\theta_{sa}=\beta x_{sa}+b_s+u_{sa},\qquad
b_s\sim N(0,\tau^2),\quad u_{sa}\sim N(0,\sigma^2).
\label{eq:discrepancy}
\end{equation}
Human estimates have sampling covariance $V_s$, including shared controls, so the likelihood uses $V_s+\tau^2\mathbf{1}\mathbf{1}^{\mathsf T}+\sigma^2 I$. An outer realization draws parameters and study/arm discrepancies once and shares them across affected agents. Drawing independent model errors for every agent would wrongly make behavioral uncertainty vanish with population size.

The tabulated tier, correction operators, evaluation harness, discrepancy analysis, and scale engine are implemented. A memory tier uses the last three sampled answers with a rank-preserving cell-level correction toward independent-answering marginals; it is evaluated separately. Network/exposure conditioning, persistent latent traits, and an integrated \texttt{init/validate/compare/report} workflow remain designs. Neither memory-plus-intervention composition nor large-scale endogenous network dynamics is validated. Figure~\ref{fig:architecture} summarizes the prediction, correction, execution, and discrepancy flows.

\section{Materials and evaluation protocol}
SocSci210 comprises 210 TESS experiments, split into 170 seen and 40 unseen studies, with 37 supporting held-out intervention comparisons \citep{kolluri2025,tess}.
Arechar provides 45 COVID headlines, sharing and accuracy judgments on six-point scales, and 16 countries \citep{arechar2023}.
Epstein provides 9,070 completed participants in five waves, eight intervention types, and 20 cards, equally divided into true and false headlines \citep{epstein2021}.

Condition-sensitivity $r$ correlates predicted and human cell means after centering within study--task blocks.
Sign accuracy evaluates pairwise directions on reliable human contrasts ($|z|\geq3$), with respondent-clustered standard errors and half credit for model ties.
Captured decision gain measures the human gain from the model's highest and lowest condition choices relative to random selection and the available human optimum, with reference values zero and one, respectively.
Effect MAE compares condition-minus-reference effects; coverage asks whether intervals contain human effect estimates, including their sampling error, rather than unobserved individual responses.

Criteria were committed before each corresponding held-out run or newly resolved outcome analysis; the full protocol, prior-exposure disclosures, retrospective-analysis qualifications, and appended errata are in Appendix~\ref{app:criteria}.

\section{Validity as calibration}
\paragraph{Marginal fit is insufficient.}
The G1 condition-blind pooled-human oracle has distance 0.0339 versus Jev's 0.1519 under the observed-range convention. It uses the evaluation study's own answers and is a diagnostic rather than a deployable baseline. It even scores below the cited published models, although task sets differ. Yet it has no condition variation; its centered correlation is undefined. A marginal leaderboard can therefore reward knowledge of the outcome distribution without requiring useful intervention predictions.

\paragraph{Decision value saturates.}
On G-A's test analysis, Jev sign accuracy is 0.671 \CI{0.500}{0.803} and captured gain is 0.268 \CI{0.050}{0.454}; permutation tests pass the declared information criteria. Under the same human judges, sign accuracy is 0.702 against a half-sample pilot's 0.985, and gain is 0.237 against 0.628. The stronger declared criterion is not met. From 5 to 200 simulated respondents per cell, sign accuracy ranges from 0.632 to 0.674 and gain from 0.255 to 0.358 without a rising trend (Appendix~\ref{app:extra}). More simulated respondents reduce persona-sampling noise but do not remedy the kernel's systematic errors.

Demographic persona information is correspondingly limited: a cross-validated demographic-group predictor has individual-response correlation 0.013. This finding concerns these representations and outcomes. It does not make all respondents interchangeable, nor exclude information in richer measurements.

\paragraph{A separately evaluated memory tier.}
In B2's six clean within-subject studies, three-answer self-memory plus rank-preserving marginal correction passes both coherence and marginal criteria in all six. Median correlation-pattern agreement rises from 0.01 to 0.52, approximately 65\% of human split-half structure; relative correlation magnitude rises from 0.32 to 0.43. Distant-pair coherence reaches only 40\% of human coherence. Raw self-conditioning can surrender 91\% of the independent kernel's marginal advantage over a uniform baseline, making the correction necessary. These are sequence-level results, not evidence for individual prediction or persistent network behavior.

\section{Content fidelity across 16 countries}
All three Arechar held-out parts pass the frozen item-and-person gate. In both non-US parts, sharing ranking, accuracy ranking, and truth discernment pass in all 15 countries, exceeding the required 12. Sharing correlations range from 0.56--0.83 on new countries/odd items and 0.61--0.89 on new countries/even items; accuracy correlations range from 0.73--0.87 and 0.69--0.95. Human split-half reliability is approximately 0.89--0.98 for sharing and 0.96--0.99 for accuracy. These are reliability references, not a guarantee about every content domain.

The nine translated-questionnaire countries perform comparably to the English group: new-country median sharing/accuracy correlations are 0.736/0.852 versus 0.711/0.832. Because the model reads English in both groups, this tests transfer across participant populations and translated human instruments, not Jev's multilingual inference. On US calibration, model accuracy discernment is 1.27 versus people's 1.49, while sharing discernment is 0.42 versus 0.57: both use only part of the truth discrimination visible in explicit accuracy judgments.

\begin{table}[tp]
\centering\small
\caption{Arechar test evidence. Upper panel: kernel gates, with prompt/tips effects in six-point units. Lower panel: models on the same 30-person-per-country-condition subset in the both-new part; intervals come from C3b's paired analysis. Human references use all eligible respondents.}
\label{tab:arechar}
\begin{tabular}{lccc}
\toprule
Test part & Content/person gate & Prompt & Tips\\
\midrule
US, even headlines & Pass & $-0.102$ (fail) & $+0.056$ (fail)\\
15 countries, odd & Pass (15/15) & $-0.034$ (fail) & $+0.005$ (fail)\\
15 countries, even & Pass (15/15) & $-0.060$ (fail) & $-0.003$ (fail)\\
\midrule
Matched-subset comparison & & Effect [95\% CI] & Effect [95\% CI]\\
Jev & & $-0.025$ [$-0.058$, 0.011] & 0.015 [$-0.022$, 0.053]\\
Astra & & 0.052 [0.010, 0.094] & 0.045 [0.002, 0.088]\\
Human reference & & 0.127 [0.089, 0.165] & 0.075 [0.036, 0.110]\\
\bottomrule
\end{tabular}
\end{table}

Intervention sensitivity is a separate capability (Table~\ref{tab:arechar}). The kernel fails both gates. Astra passes both, with a prompt advantage over Jev of 0.077 \CI{0.055}{0.099} and a tips advantage of 0.030 \CI{0.008}{0.051}. Astra exceeds Jev's prompt effect in 13/15 countries; this is not 13/15 agreement with human directions. Its pooled effects reach approximately 40--60\% of human effects. Tips evidence is marginal under multiplicity correction: its one-sided $p=0.026$ passes the declared per-intervention rule but not a two-test Bonferroni threshold.

Cross-country comparisons are not supported. Descriptively, predicted accuracy discernment spans only 1.14--1.30 while human discernment spans 0.71--2.00. Nor do these familiar accuracy interventions establish inference on novel interventions. The A3c publication-status test finds no detectable interaction in flagship advantage, 0.044 \CI{-0.078}{0.170}; recall versus inference remains unresolved.

\section{Sparse flagship calibration}
\subsection{Epstein: lower intervention-effect error}
D1 applies the frozen cell rule to held-out participants, content, and intervention variants. The 73,600 kernel predictions cost \$1.60. Astra supplies 1,080 paired anchor predictions, approximately 1.7\% of the 64,000 application states, plus 9,000 predictions on a disjoint direct-audit panel. All 60 cell corrections are feasible. Anchor and audit workloads together use 4.10 million Codex tokens.

\begin{table}[tp]
\centering\small
\caption{Epstein's declared secondary effect endpoint and subsequent robustness analyses. Raw MAE intervals resample participants within wave and arm. Calibrated MAE fits a slope on other waves and predicts the excluded wave; it is a post-run robustness analysis. Model-only RMSE subtracts estimated human sampling variance from observed squared error.}
\label{tab:epstein}
\begin{tabular}{lccc}
\toprule
System & Raw effect MAE [95\% CI] & Calibrated MAE & Model-only RMSE\\
\midrule
Kernel & 0.0305 [0.0242, 0.0412] & 0.0376 & 0.033\\
Hybrid & 0.0180 [0.0132, 0.0289] & 0.0183 & 0.014\\
Flagship direct audit & 0.0200 [0.0153, 0.0314] & 0.0326 & 0.020\\
\bottomrule
\end{tabular}
\end{table}

Effect MAE falls by 41\% (Table~\ref{tab:epstein}). The paired absolute-error improvement is 0.0125, with participant-bootstrap interval \CI{0.0064}{0.0167} and effect-bootstrap interval \CI{0.0025}{0.0222}; a proportion below 0.001 of participant resamples give a nonpositive improvement. Leaving out each effect-bearing wave gives reductions of 38--44\%. Out-of-wave calibration retains lower hybrid error than either calibrated parent. Kernel slopes are near zero or negative, whereas hybrid slopes remain positive. These observations suggest that the correction supplies useful effect information rather than merely rescaling the kernel.

The human effects' mean standard error is 0.0147, implying an approximate perfect-model MAE floor of 0.0117. The hybrid's estimated model-only RMSE, 0.014, is comparable to that sampling uncertainty; its observed MAE of 0.0180 is not equal to the floor. With the same anchors, the persona-count curve is essentially flat from 10 to 200 per wave: kernel MAE is about 0.0305--0.0308 and hybrid MAE about 0.0180--0.0181. The available error budget is better spent on correction than on expanding this simulated population.

\begin{figure}[tp]
\centering\includegraphics[width=\linewidth]{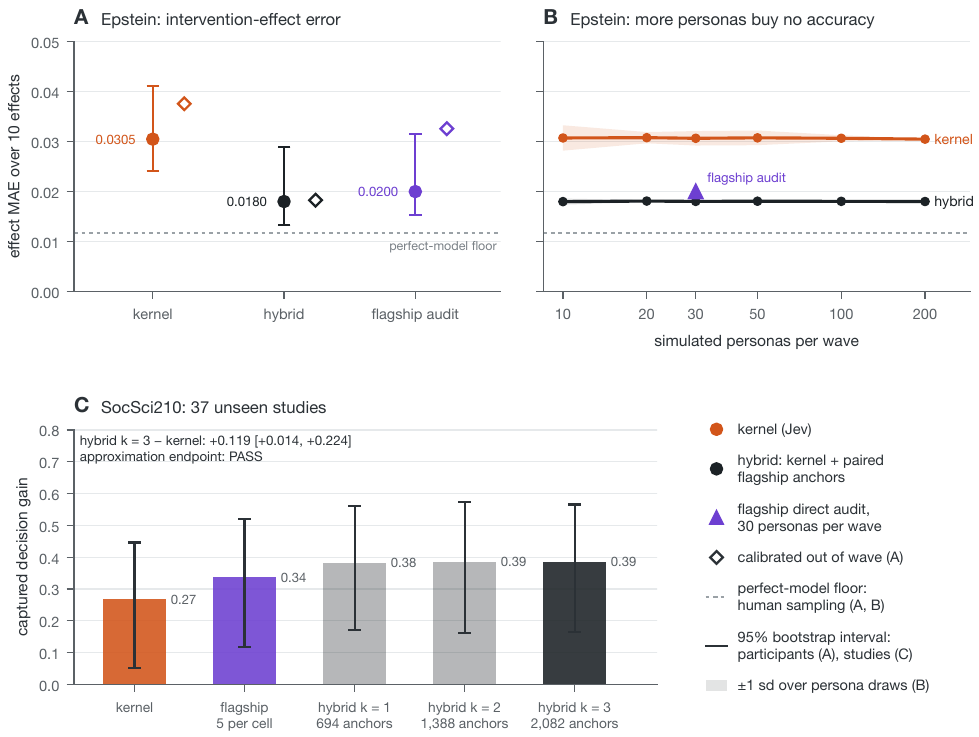}
\caption{Sparse flagship calibration is shown across two datasets. Top row: (A) Epstein raw effect MAE, with values beside points and 95\% participant-bootstrap intervals; hollow diamonds show out-of-wave calibrated MAE and the dotted line marks the perfect-model sampling floor (0.0117). (B) Mean MAE versus personas per wave with the same 1,080 anchors, A's vertical scale, and directly labeled lines; bands show $\pm1$ persona-subsampling standard deviation, not human uncertainty, and the triangle shows the direct audit. Bottom row: (C) SocSci210 captured gain with values beside bars, 95\% study-bootstrap intervals, and 694/1,388/2,082 paired anchor predictions for $k=1/2/3$ below the bars, alongside kernel and five-per-cell flagship references; the key is to the right. Epstein effect MAE is secondary; SocSci210's primary approximation endpoint passes narrowly.}
\label{fig:calibration}
\end{figure}

The preregistered primary policy-value gain was not shown: hybrid minus kernel is 0.0027 \CI{-0.0118}{0.0169}. The best-minus-worst policy range is only 0.039 across five decisions, and the frozen power note anticipated limited resolution. Same-judge values are 0.0938 for the hybrid, 0.0912 for the kernel, 0.0875 for direct audit, and 0.0941 for a human half-sample pilot; this numerical proximity is not an equivalence test.

The per-wave choices reveal both benefit and imported error (Appendix~\ref{app:epstein}). In wave 3, Jev selects generic norms, the human worst arm, predicting $+0.032$ against $-0.011$ observed; the hybrid selects evaluation, the second-best arm. Jev also overpredicts wave-4 partisan norms ($+0.056$ versus $+0.006$). In wave 5, the kernel correctly selects importance-plus-norms, but replacement imports Astra's preference for importance. Correction is fallible even when its average error improves.

Headline heterogeneity, declared separately before its computation, is not established: hybrid within-arm $r=0.157$ \CI{-0.001}{0.241}. Direct audit yields 0.392 \CI{0.178}{0.433}, and kernel control sharing alone yields 0.383 \CI{0.158}{0.433}, against human split-half reliability 0.392. Tilting treated predictions retains the kernel's mistaken within-cell responses. A control-distribution tilt is a candidate for a future frozen test, not a validated repair here.

\subsection{SocSci210: improved captured decision gain}
D3 queries three paired anchors per condition in 162 blocks from 37 held-out studies: 2,082 items, 1.00 million tokens, and 15 minutes. The reference is the earlier unpaired five-per-cell flagship run of 3,615 items. The primary approximation correlation rises from 0.608 to 0.709 \CI{0.550}{0.832}; its paired gain is 0.101 \CI{0.003}{0.293}. Both frozen requirements are met, but the margin is narrow. This is evidence that the hybrid moves toward the flagship, not that it reproduces the flagship exactly.

The strongest human-referenced secondary is captured gain: 0.268 to 0.386, an improvement of 0.119 \CI{0.014}{0.224}. One anchor per condition, 694 items or 0.5\% of population states, already yields a gain of 0.114 \CI{0.014}{0.215}; two yield 0.118 \CI{0.006}{0.224}. Three cover 1.5\%. The direct flagship's gain is 0.336, with improvement over the kernel 0.069 \CI{-0.037}{0.168}. We do not claim hybrid superiority to the flagship: their gain difference is 0.050 with declared 90\% interval \CI{-0.018}{0.117}.

Sign accuracy rises from 0.671 to 0.709 at three anchors, difference 0.038 \CI{-0.044}{0.122}. Within-task $r$ changes little, 0.397 to 0.423 (difference 0.026, \CI{-0.120}{0.213}), and raw effect MAE worsens from 0.052 to 0.083 (difference 0.031, \CI{0.016}{0.047}). The hybrid inherits overreaction: magnitude ratio 1.75 versus the kernel's 0.68. Applying frozen development slopes gives MAE 0.056/0.052/0.055 for hybrid/kernel/flagship. Better choices therefore coexist with worse uncalibrated magnitudes.

Pairing is not shown to outperform equal-budget unpaired prediction: three-anchor differences are sign accuracy 0.012 \CI{-0.054}{0.072}, gain 0.047 \CI{-0.062}{0.148}, and $r=-0.038$ \CI{-0.103}{0.054}. D1's descriptive paired-anchor/direct-audit comparison does not isolate pairing either. The supported result is access to comparable flagship decision quality through sparse calls, not a general advantage of paired anchoring.

On reliable test contrasts, kernel and hybrid agree on 78.9\%, with accuracy 0.742 there. On disagreements, the hybrid scores 0.586 and the kernel 0.406. This is weaker than the development agreement diagnostic and is not a calibrated runtime confidence rule. Appendix~\ref{app:d3} reports every system and the declared comparisons. Epstein and SocSci210 are the only independent human-referenced tests of the hybrid.

\section{Discrepancy propagation}
D2 fits Equation~\ref{eq:discrepancy} on 91 seen studies and 2,195 effects for retrospective evaluation on 37 held-out studies, using maximum likelihood with human sampling covariance and 80 study-bootstrap fits. For Jev, $\beta=0.925$ \CI{0.61}{1.17}, $\tau=0.037$ \CI{0.021}{0.046}, and $\sigma=0.080$ \CI{0.051}{0.128}. Split-half coverage is 0.58/0.81/0.88 at nominal 0.50/0.80/0.90. Retrospective test coverage is 0.67/0.93/0.96; human-sampling-only coverage is 0.14/0.29/0.36. At 90\%, mean width is 0.324 and interval score 0.422, compared with 0.064 and 0.746 without discrepancy (lower score is better).

\begin{figure}[tp]
\centering\includegraphics[width=\linewidth]{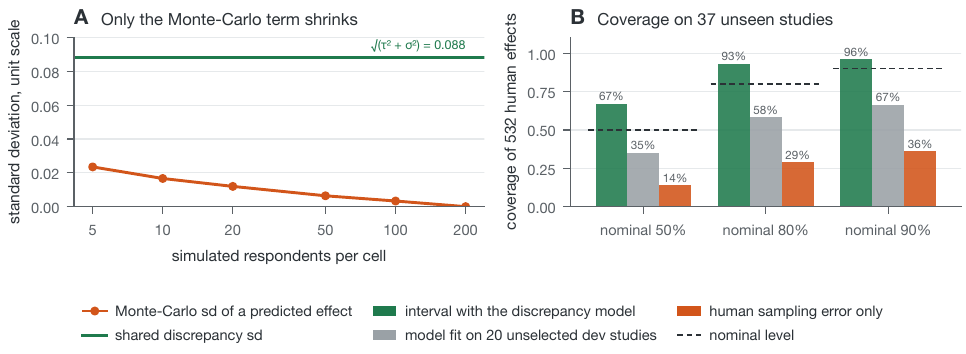}
\caption{Why additional simulated respondents do not eliminate behavioral uncertainty. A shared legend appears below the panels. (A) Effect standard deviation from persona subsampling at 5--200 respondents per cell, compared with the shared discrepancy scale $\sqrt{\tau^2+\sigma^2}=0.088$. The endpoint at 200 is zero only relative to the finite table. (B) Retrospective, study-weighted test coverage at nominal 50\%, 80\%, and 90\% (dashed lines), comparing discrepancy fitted on 91 selected studies, discrepancy fitted on 20 unselected development studies, and human sampling error alone. Discrepancy is shared, not drawn independently per agent.}
\label{fig:discrepancy}
\end{figure}

The discrepancy scale, $\sqrt{\tau^2+\sigma^2}=0.088$, exceeds persona-sampling standard deviation 0.0235 at five respondents per cell and 0.0033 at 100 (Figure~\ref{fig:discrepancy}). Astra also needs discrepancy: $\beta=0.579$ \CI{0.32}{0.66}, with test coverage 0.63/0.86/0.94. Its predictions cannot serve as human truth simply because they cost more.

The test intervals are conservative, plausibly because fitting studies were selected for reliable contrasts. Jev's slope changes to 0.661 on 20 unselected development studies and approximately 0.50 on test reliable contrasts; magnitude calibration is unstable. Current residual estimates also depend on simulation sample size. Explicitly separating excess discrepancy from Monte Carlo variance and refitting on an unselected, deployment-matched set are designed improvements, not completed validations. Transferring the D2 interval to Epstein covers 10/10 effects for kernel and hybrid, but half-width 0.145 exceeds the largest human effect, 0.062. That prospective check is uninformative about transport calibration.

\section{Cost and scale}
\paragraph{Quality per expensive prediction.}
On the same intended 3,615 SocSci210 items, Jev, Luna, and Astra marginal distances are 0.136/0.132/0.101; captured gains are 0.272/0.305/0.336 (Table~\ref{app:models}). GPT-5.6 Luna is the cheapest comparison tier, not a flagship. Astra's gain advantage over Jev is 0.064 \CI{-0.048}{0.168}. All three err together on 19.4\% \CI{9.6\%}{32.9\%} of reliable contrasts, versus 3.5\% under error independence. Higher distributional fit does not establish a uniform decision advantage.

The Codex CLI reports total tokens per call, including its wrapper, without an input/output split. We assume 70\% input and 30\% output tokens and compute cost as total tokens times $(0.7p_{\mathrm{in}}+0.3p_{\mathrm{out}})/10^6$, where prices are per million tokens. OpenAI list prices read on September 21, 2026 were Luna \$0.20/\$1.20, Terra \$2/\$12, Sol \$4/\$20, and Astra \$10/\$50 for input/output, giving blended prices of \$0.50, \$5.00, \$8.80, and \$22.00.\footnote{The author archived the price page on September 21, 2026.} Codex usage was billed through a subscription, not an API invoice.

For SocSci210, A3 (GPT-5.6 Luna) used 1,955,016 tokens for 3,615 predictions (541 tokens per prediction), implying \$0.27 per thousand predictions; A3b (GPT-6 Astra) used 2,570,752 for 3,615 (711), implying \$15.65. C3b (Astra, Arechar both-new) used 5,809,259 for 12,833 (453), implying \$9.96 per thousand. D1 (Astra, Epstein anchors plus audit) used 4,098,094 for 10,080 (407), implying \$8.94 per thousand and approximately \$90 total; the 1,080 anchors alone account for approximately 0.44 million tokens and \$9.7. D3 (Astra, SocSci210 paired anchors) used 1,004,653 for 2,082 (483), implying \$10.62 per thousand and approximately \$22 total. Running the flagship over the full SocSci210 table is extrapolated at 66--97 million tokens.

Measured Jev costs per thousand predictions are \$0.0344 in the A3-era SocSci210 run, \$0.0331 for the full SocSci210 test table (\$4.53 for 136,901 predictions), \$0.0319 for Arechar held-out, \$0.0217 for Epstein D1, and \$0.0227 for fresh S1 predictions. The implied cost ratios are approximately 8 times Jev for Luna (A3), 455 times for Astra (A3b, using the A3-era Jev price), and 312 times for Astra on Arechar (C3b).

\paragraph{Prediction and execution measurements.}
S1 uses 500 previously unused Epstein wave-3 personas, 20 cards, and control/tips conditions. The 20,000-state workload completes in 920.7\,s, with 1,303 predictions per minute including reuse and approximately 922 live calls per minute under a 960-call throttle. Median latency is 213\,ms, p95 304\,ms. The \$0.322 total implies \$0.0227 per thousand fresh live predictions; 29.2\% of requested states are exact cache hits. One Cloudflare 520 is retried; no malformed distributions remain.

\begin{table}[tp]
\centering\small
\caption{Execution from the fixed S1 table on a laptop. Each step evaluates both conditions; reported agent-steps follow the source record's convention. These are timings for tabulated fixed-exposure decisions, not fresh inference or network simulation.}
\label{tab:scale}
\begin{tabular}{lrrr}
\toprule
Engine & Agents & Seconds for 20 steps & Agent-steps/s\\
\midrule
Vectorized & $10^4$ & 0.007 & $2.7\times10^7$\\
Vectorized & $10^5$ & 0.075 & $2.7\times10^7$\\
Vectorized & $10^6$ & 0.90 & $2.2\times10^7$\\
Scalar & $10^4$ & 0.95 & $2.1\times10^5$\\
\bottomrule
\end{tabular}
\end{table}

Execution cycles these 500 personas through the table (Table~\ref{tab:scale}); it does not create a million distinct representations. Peak resident memory is 685\,MB. Across 100 seeds at $10^4$ agents, common random numbers reduce effect standard deviation from 0.00303 to 0.00072, approximately an 18-fold variance reduction. The exact conditional effect is 0.0221. Reusing D1's recorded shifts requires no new flagship calls and makes all six cells feasible, yielding 0.0065. This demonstrates operational reuse, not new human validation: the source wave's tips effect was underestimated by the hybrid. The S1 run and all execution timings used one Apple M4 Pro laptop (14 cores, 48\,GB RAM), macOS 26.6.2, Python 3.12.2, NumPy 2.5.3, and pandas 3.0.6, in a single process. Wall-clock timings use \texttt{time.perf\_counter} around the execution loop only, excluding table construction; peak resident memory comes from \texttt{resource.getrusage}.

These measurements depend on proprietary model access and recorded versions; reproducible code does not guarantee regeneration of vendor outputs.

\section{Licensed uses and evidence records}
An evidence record states the property being reported, its reference population and stimuli, model and configuration hashes, operator provenance, uncertainty method, and unsupported assumptions. It separates conditional-model estimates, execution noise, and human-discrepancy intervals. Correction records also expose requested and achieved means, feasibility, affected weight, and cells where the original kernel effect was replaced. Existing D1 records instantiate this structure; the integrated user-facing report remains a design.

\begin{table}[tp]
\centering\small
\caption{Property-specific evidence record for a misinformation experiment. A supported property does not transfer automatically to another row.}
\label{tab:evidence}
\begin{tabularx}{\linewidth}{@{}p{0.27\linewidth}X@{}}
\toprule
Property & Evidence and permitted interpretation\\
\midrule
Content ranking & Kernel passes Arechar test gates; screening within the evaluated domain is supported.\\
Known intervention direction & Sparse flagship calibration improves Epstein effect error and SocSci210 decision gain; retain endpoint and domain qualifications.\\
Human effect intervals & D2 improves retrospective SocSci210 coverage; transport to the hybrid or another domain is not established.\\
Cost--quality comparison & Measured calls, tokens, table execution, and declared human metrics support comparison on these workloads.\\
Individual/country effects & Individual prediction and cross-country susceptibility comparisons are unsupported.\\
Repeated exposure/network & Conditional scenarios only; cascade sizes and endogenous network responses are unvalidated.\\
\bottomrule
\end{tabularx}
\end{table}

The supported uses are content screening, direction screening for literature-known intervention families with calibration, and cost--quality comparison with explicit discrepancy. A narrow interval from repeated simulation supplies no additional license. Neither agreement between models nor a provider's confidence field is a calibrated probability that a conclusion about people is correct.

\section{Limitations}
Only two held-out human-referenced datasets test the hybrid: Epstein and SocSci210; their effect structures and prior exposure differ. D1's policy-value primary and the later headline-heterogeneity primary were not shown, and only ten wave-relative effects support its error result. Numerical proximity to a half-sample human pilot establishes no equivalence.

D3's primary approximation gain passes narrowly; within-task sensitivity is unchanged within uncertainty and raw effect MAE worsens. Pairing does not demonstrably outperform equal-budget unpaired prediction. Sparse calibration can import flagship errors, as in Epstein wave 5, while the kernel itself overpredicts norms messages. Neither parent is a universally better behavioral model.

Effect magnitudes require further calibration: discrepancy slopes vary across study selections, and the broad Epstein coverage check does not validate cross-domain uncertainty. A3c is inconclusive about recall; familiar experimental designs may be recognized from training data. The experiments therefore do not license literature-absent intervention prediction.

Memory coherence is local, and cohort-level marginal correction may erase real persistence or treatment pathways when exposure is endogenous. An offline rank mapping also makes an agent's corrected answer depend on other agents in its cell. Large tabulated execution does not validate that mapping as an online transition process, network diffusion, or cascade scale.

Stimulus reconstruction, omitted card images, representative within-cell stimuli, demographic compression, and supported-scale exclusions constrain fidelity. Proprietary models and access limits constrain replication: reproducible code does not imply reproducible outputs. Version pins and private caches preserve recorded calculations, but vendor updates, repeat-call variability, and unavailable response caches prevent guaranteed independent regeneration of the same predictions.

\section{Ethics, terms, and availability}
The demonstrations concern harm reduction through misinformation interventions. They do not demonstrate persuasion optimization, election forecasting, or replacement of human participants. Simulated demographic profiles may reproduce stereotypes; content-ranking evidence must not be used to infer an individual's beliefs or rank national populations by susceptibility. Personally identifying fields in the source exports are excluded from the modeled state.

Code and evaluation records are available at \url{https://github.com/HengyuLi-Ozaki-lab/kite_population_simulator} (MIT license): the \texttt{kite} package, the frozen criteria files with their appended errata, the aggregate results and provenance records. Only aggregates are redistributed; the Jev response cache is not released and outputs are not used for distillation. The Python package and CLI are both named \texttt{kite}. SocSci210 is obtained from \texttt{socratesft/SocSci210}; its dataset license is unstated in the audit and original-source terms remain applicable. Arechar's OSF project \texttt{g65qu} likewise states no license in the audit; its microdata are not redistributed. Epstein's Dataverse dataset (\href{https://doi.org/10.7910/DVN/18SHLJ}{doi:10.7910/DVN/18SHLJ}) is CC0; the article is CC BY 4.0 and the separate OSF stimulus files have no stated license. We do not treat a data license as licensing every stimulus asset.

Under the Jev terms recorded in the project audit, outputs are not used for distillation or to train an imitation model, and the full response cache is not released. Correction operates on task-specific distributions and does not train a surrogate.

\paragraph{Declarations.}
This work is personal research with no funding of any kind and no competing interests. The study analyzed only publicly available, de-identified datasets (SocSci210; Arechar et al.\ 2023 materials; Epstein et al.\ 2021 data) and collected no new data from human participants. Beyond the models evaluated in this paper, generative AI tools were used throughout the project: Claude (Anthropic; Fable~5.1 and Opus~5.5 models, via Claude Code) for analysis code and evaluation records. The author directed the work and takes full responsibility for all content.

\clearpage
\appendix
\section{Frozen criteria, prior exposure, and errata}\label{app:criteria}
\subsection{Datasets, sampling, and stimuli}\label{app:materials}
\paragraph{SocSci210.}
Scale support and filtering leave 39 studies for the marginal analysis, and shared-scale comparisons leave 37 studies, 162 blocks, 532 reference-condition effects, and 304 reliable pairwise contrasts for D3. The kernel run contains 136,901 predictions, with up to 200 respondent rows per cell; this is the full \emph{simulation table}, not every original human respondent. Human reference distributions use all eligible rows. The prepared marginal reference has 456,587 response rows, not that many distinct people.

Development for D3 and D2 uses 91 seen studies, 236 comparable blocks, and 2,195 effects, selected through the presence of reliable contrasts. The earlier unselected development set contains 20 studies. These sets answer different questions and must not be pooled as independent replications. D3 anchors are the first three respondents in deterministic hash order, each queried under every condition; smaller budgets are nested. One representative stimulus is used per condition, although 24\% of cells contain stimulus variants.

\paragraph{Arechar.}
Accuracy prompts have an established experimental literature \citep{pennycook2021}. Calibration used the United States and odd-indexed items. The three test parts are US/even, other countries/odd, and other countries/even. Models read US English texts throughout; questionnaire language is a declared moderator. Personas include pretreatment social-media information and demographic attributes treated as unaffected by the prompt, excluding post-treatment attitudes.

\paragraph{Epstein.}
Randomization was within wave; ten wave-relative sharing effects result. The kernel uses 200 application and 30 disjoint audit personas per wave. Models receive transcribed source/headline text and intervention screens; original cards included images. Long Evaluation is described because its underlying image sequence is unavailable. Published pooled effects were seen during stimulus transcription, but scored wave-by-arm outcomes were unopened until policies were locked.

\subsection{Commitment, prior exposure, and errata}
Criteria files specify representations, seeds, endpoints, nulls, and verdicts before the corresponding held-out run or newly resolved outcome analysis. Loaders and scripts enforce commitment of the relevant configuration. Table~\ref{tab:criteria} records the freezes and prior exposure for each analysis. These are version-controlled preregistrations with differing prior exposure, not a claim that all test information was unseen; D2's test coverage is retrospective. Original criteria remain intact; corrections are appended below.

\begin{table}[!htbp]
\centering\footnotesize
\caption{Criteria and their evidential status. Paths are relative to \texttt{materials/configs/}. A freeze concerns a specified analysis or run, not every earlier use of the dataset.}\label{tab:criteria}
\begin{tabularx}{\linewidth}{@{}p{0.16\linewidth}p{0.34\linewidth}X@{}}
\toprule
Record & Frozen source & Main requirement and status\\
\midrule
G1 & \path{frozen.yaml} & p3/choice selected on development, commit \texttt{389b984}; marginal and condition gates pass. Published-model comparisons require metric alignment.\\
G-A & \path{decision_value.yaml} & Sign: permutation $p<0.01$ and above null p95; gain: $p<0.01$ and bootstrap interval above zero. Both pass; stronger two-thirds-of-pilot criteria do not. Earlier test sensitivity was known.\\
C2/C3 & \path{arechar_split.yaml}, \path{arechar_criteria.yaml} & Split fixed before download; criteria committed at \texttt{d826501} after US/odd calibration. Content/person gates pass; kernel intervention gates fail.\\
C3b/A3b & \path{flagship_baseline.yaml} & Flagship comparison frozen at \texttt{ffcb4a7} before predictions; existing criteria reused unchanged.\\
D1 & \path{epstein_criteria.yaml}, \path{epstein_screens.yaml} & Commit \texttt{7b5a5a3}; policies and prediction hashes locked before scoring. Positive policy-value difference with interval excluding zero required. Not shown. Effect MAE remains secondary.\\
D1 A2 & \path{epstein_headline_criteria.yaml} & Commit \texttt{cdc7864}, after arm aggregates but before headline outcomes were examined; positive within-arm correlation with interval above zero required. Not shown.\\
D3 & \path{socsci210_paired_anchors.yaml} & Commit \texttt{c701187} before test anchors. Three-anchor approximation $r\geq0.60$ and positive gain with 95\% interval above zero required. Pass. Earlier kernel/flagship test summaries and model-only agreement were known.\\
B2 & \path{b2_criteria.yaml} & Commit \texttt{3374f3f} before predictions. Coherence and marginal gates in at least 5/6 studies required; achieved 6/6 on both.\\
A3c & \path{publication_moderator.yaml} & Freeze before publication coding and seen-study flagship predictions. Positive interaction with permutation $p<0.05$ required; not detected.\\
D2 & Analysis specification in D2 record & Maximum-likelihood discrepancy fit and retrospective coverage; no supplied standalone frozen D2 criterion.\\
\bottomrule
\end{tabularx}
\end{table}

D1's appended errata preserve the original text: (i) ``13 effects'' is a counting error; the sharing design yields ten and the analysis enumerated these; (ii) the forecast of roughly 11,000 flagship items and \$2.3--2.5 kernel cost is replaced by actual 10,080 items and \$1.60; (iii) scoring ran twice because a Path-object serialization error prevented the initial report from being written or printed, followed by a serialization-only fix and the same policies and seed; (iv) the broad D2 coverage interval is recorded as uninformative, without changing the endpoint. D1 headline criteria and D3 report no errata. G-A likewise documents a date-object serialization rerun after results were printed, with unchanged analysis and seed.

For Arechar, item correlations must exceed item-shuffle p95 with permutation $p<0.01$, in at least 12/15 countries for each non-US part. Truth discernment must exceed a truth-shuffle null, and within-item person discrimination must exceed its people-shuffle p95. The intervention rule requires both a model effect above condition-shuffle p95 and a human 95\% interval above zero; it is applied per treatment without multiplicity adjustment.

For B2, magnitude and structure must exceed independent-answer null p95 based on 40 seeds; corrected marginal distance must be no greater than the redraw-floor p95 based on 200 repetitions. Selection excludes repeated task identifiers within respondents and previously used individual-level studies. In A3c, two independent agent coders, blind to model results, coded publication status; reconciliation preceded analysis. Results-public status is a proxy for possible exposure, not a verified model-training manifest.

\subsection{Metric definitions and uncertainty}\label{app:metrics}
The records' \emph{L1 validity level} is marginal distribution fit, measured by Wasserstein distance on the stated $[0,1]$ answer scale: $W_1(P,Q)=\int_0^1|F_P(t)-F_Q(t)|\,dt$. It is not the L1 distance between probability vectors. Distances are averaged within studies and then across studies. The separately reported published convention uses observed response ranges; it is not interchangeable with stated-scale normalization.

For each study--task block with at least three conditions and finite means, captured gain is $(\overline h_{\arg\max m}-\overline h)+(\overline h-\overline h_{\arg\min m})$ and available gain is $\max h-\min h$, where $h$ and $m$ are human and model condition means and ties average human values over the tied conditions; other blocks contribute zero to both sums, the reported ratio $\sum\mathrm{captured}/\sum\mathrm{available}$ weights blocks by available gain, and the study bootstrap reweights blocks by resampled study. Same-judge pilots select on one human half and are evaluated alongside models on the other; reliability screening uses the judging half.

For misinformation, discernment is mean sharing of true minus false headlines and the effect is its treatment-minus-control change; Arechar results retain six-point units, Epstein uses binary sharing probabilities.

Unless specified otherwise, intervals are 95\% bootstrap intervals. D1 resamples participants within wave and arm; D3 resamples studies, preserving dependent effects; D2 uses study-bootstrap parameter draws and human sampling covariance. D3 comparisons to the five-per-cell flagship use the declared 90\% intervals. Monte Carlo precision is reported separately from uncertainty over studies or participants.

\section{Epstein: choices, effects, and robustness}\label{app:epstein}
\begin{table}[!htbp]
\centering\small
\caption{Locked within-wave choices and observed discernment of the chosen arm. E = evaluation; LE = long evaluation; GN = generic norms; TN = tips plus norms; I = importance; IN = importance plus norms. Wave 1 has no sharing intervention. Human choices are descriptive optima using the full sample. Source: D1 decision record and \texttt{d1-policies-summary.json}.}
\begin{tabular}{cllll}
\toprule
Wave & Human best & Kernel choice (human value) & Hybrid choice (human value) & Audit choice\\
\midrule
1 & Control & Control (0.059) & Control (0.059) & Control\\
2 & LE & LE (0.128) & LE (0.128) & LE\\
3 & Tips & GN (0.057) & E (0.089) & GN\\
4 & TN & TN (0.113) & TN (0.113) & TN\\
5 & IN & IN (0.101) & I (0.083) & I\\
\bottomrule
\end{tabular}
\end{table}
\begin{table}[!htbp]
\centering\small
\caption{Ten wave-relative discernment effects on the binary sharing scale. Human standard errors and 95\% intervals use a participant bootstrap within wave and arm (4,000 resamples, seed 0); $n_T/n_C$ gives treated/control participants. Controls are wave-specific. Human estimates come from \texttt{results/d1/human\_effect\_intervals.json}, as supplied in \texttt{RESOLUTIONS.md}; model effects retain D1 rounding.}\label{tab:human-effects}
\footnotesize\setlength{\tabcolsep}{3pt}
\begin{tabular}{@{}clrrcrrrr@{}}
\toprule
Wave & Intervention & Human & SE & 95\% interval & $n_T/n_C$ & Kernel & Hybrid & Audit\\
\midrule
2 & Evaluation & $+0.0455$ & 0.0175 & $[+0.0126, +0.0794]$ & 395/387 & $-0.013$ & 0.023 & 0.008\\
2 & Long evaluation & $+0.0624$ & 0.0172 & $[+0.0291, +0.0966]$ & 410/387 & $0.011$ & 0.025 & 0.013\\
3 & Evaluation & $+0.0202$ & 0.0143 & $[-0.0072, +0.0483]$ & 540/533 & $-0.015$ & 0.011 & 0.015\\
3 & Generic norms & $-0.0114$ & 0.0139 & $[-0.0379, +0.0153]$ & 510/533 & $0.032$ & 0.008 & 0.017\\
3 & Tips & $+0.0265$ & 0.0149 & $[-0.0019, +0.0557]$ & 498/533 & $0.023$ & 0.006 & 0.016\\
4 & Partisan norms & $+0.0058$ & 0.0128 & $[-0.0198, +0.0302]$ & 949/487 & $0.056$ & 0.024 & 0.027\\
4 & Tips & $+0.0335$ & 0.0159 & $[+0.0017, +0.0651]$ & 408/487 & $0.020$ & 0.024 & 0.020\\
4 & Tips plus norms & $+0.0458$ & 0.0134 & $[+0.0199, +0.0719]$ & 934/487 & $0.060$ & 0.032 & 0.042\\
5 & Importance & $+0.0325$ & 0.0133 & $[+0.0062, +0.0581]$ & 1,046/498 & $0.019$ & 0.028 & 0.027\\
5 & Importance plus norms & $+0.0507$ & 0.0134 & $[+0.0234, +0.0761]$ & 1,072/498 & $0.029$ & 0.027 & 0.027\\
\bottomrule
\end{tabular}
\end{table}
\begin{table}[!htbp]
\centering\small
\caption{D1 policy-value results. Intervals for differences use 2,000 participant bootstraps. Chance references use 5,000 within-wave permutations; same-judge values use 200 splits. No system exceeds its chance baseline at $p<0.05$.}
\begin{tabular}{lrrrr}
\toprule
System & Policy value & Chance mean/p95 & One-sided $p$ & Same judge\\
\midrule
Kernel & 0.0915 & 0.0808/0.0944 & 0.105 & 0.0912\\
Hybrid & 0.0942 & 0.0809/0.0944 & 0.062 & 0.0938\\
Direct audit & 0.0878 & 0.0807/0.0942 & 0.197 & 0.0875\\
Human half-sample & --- & --- & --- & 0.0941\\
\midrule
\multicolumn{3}{l}{Hybrid minus kernel (primary)} & \multicolumn{2}{r}{0.0027 [$-0.0118$, 0.0169]}\\
\multicolumn{3}{l}{Hybrid minus direct audit} & \multicolumn{2}{r}{0.0063 [$-0.0079$, 0.0211]}\\
\multicolumn{3}{l}{Direct audit minus kernel} & \multicolumn{2}{r}{$-0.0036$ [$-0.0184$, 0.0112]}\\
\bottomrule
\end{tabular}
\end{table}

\begin{table}[!htbp]
\centering\small
\caption{Persona-count robustness with fixed anchors. Entries at fewer than 200 personas are means $\pm$ standard deviations across 20 subsamples, not confidence intervals; the full-table row is deterministic conditional on recorded predictions. Source: \texttt{robustness.json}.}
\begin{tabular}{rrr}
\toprule
Personas per wave & Kernel MAE & Hybrid MAE\\
\midrule
10 & $0.0307\pm0.0025$ & $0.0180\pm0.0006$\\
20 & $0.0308\pm0.0011$ & $0.0181\pm0.0002$\\
30 & $0.0307\pm0.0014$ & $0.0180\pm0.0003$\\
50 & $0.0308\pm0.0015$ & $0.0181\pm0.0002$\\
100 & $0.0307\pm0.0006$ & $0.0180\pm0.0001$\\
200 & 0.0305 & 0.0180\\
\bottomrule
\end{tabular}
\end{table}
The true/false channel MAEs are kernel 0.0139/0.0289, hybrid 0.0235/0.0224, and direct audit 0.0222/0.0231. Thus reduced discernment error combines improved false-card response with worse true-card response. The descriptive effect correlations with people are 0.60 for paired calibration and 0.05 for direct audit; their different panels and prediction budgets preclude identifying a causal benefit of pairing.

\begin{figure}[!htbp]
\centering\includegraphics[width=\linewidth]{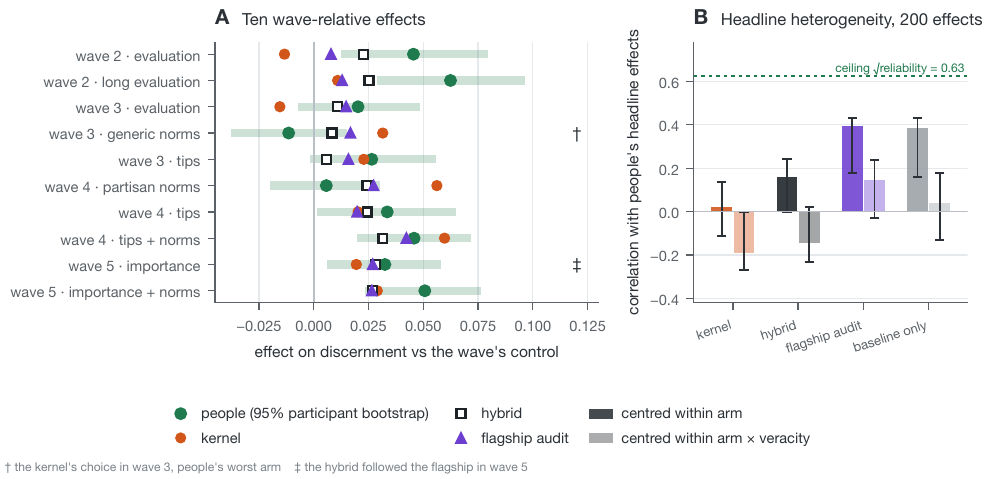}
\caption{Human and model effects by wave and arm. A shared legend appears below the panels. (A) Ten discernment effects relative to wave-specific controls: human estimates (green circles), kernel (orange circles), hybrid (open squares), and direct audit (purple triangles). Human bars are the per-effect 95\% participant-bootstrap intervals from Table~\ref{tab:human-effects}, computed within wave and arm with 4,000 resamples and seed 0. Annotations identify the kernel's wave-3 generic-norms error and the flagship choice imported in wave 5. (B) Headline-effect correlations centered within arm (dark bars) and within arm and veracity (light bars), with 95\% participant-bootstrap intervals; the baseline uses control sharing probabilities alone. The dotted line is the square root of human split-half reliability, approximately 0.63, not an uncertainty interval.}
\label{fig:epstein-detail}
\end{figure}

\section{SocSci210: full paired-anchor results}\label{app:d3}
\begin{table}[!htbp]
\centering\small
\caption{D3 test results from \texttt{report.json}: 37 studies, 162 blocks, 532 effects, 304 reliable contrasts. $F_k$ uses $k$ unpaired flagship respondents per cell; $H_k$ uses $k$ paired anchors per condition plus the kernel table. $r_F$ is correlation of effects with $F_5$. Calibration uses development slopes only.}
\begin{tabular}{lrrrrrr}
\toprule
System & Within-task $r$ & Sign & Gain & Raw MAE & Magnitude ratio & $r_F$\\
\midrule
Kernel & 0.397 & 0.671 & 0.268 & 0.052 & 0.68 & 0.608\\
$F_5$ & 0.489 & 0.671 & 0.336 & 0.078 & 1.64 & 1.000\\
$F_3$ & 0.461 & 0.697 & 0.339 & 0.086 & 1.87 & 0.927\\
$H_1$ (694 items) & 0.433 & 0.717 & 0.382 & 0.086 & 1.85 & 0.701\\
$H_2$ (1,388 items) & 0.417 & 0.742 & 0.386 & 0.084 & 1.80 & 0.721\\
$H_3$ (2,082 items) & 0.423 & 0.709 & 0.386 & 0.083 & 1.75 & 0.709\\
\midrule
\multicolumn{7}{l}{Development-scaled MAE: kernel 0.052; $F_5$ 0.055; $H_3$ 0.056.}\\
\bottomrule
\end{tabular}
\end{table}
\begin{table}[!htbp]
\centering\footnotesize
\caption{D3 test 95\% study-bootstrap intervals for each system. Magnitude-ratio intervals are not supplied. $F_3$ and $F_5$ share respondents, inflating their approximation correlation.}
\begin{tabular}{lccccc}
\toprule
System & Within-task $r$ & Sign accuracy & Captured gain & Effect MAE & $r_F$\\
\midrule
Kernel & [0.061, 0.613] & [0.495, 0.801] & [0.051, 0.446] & [0.041, 0.065] & [0.323, 0.780]\\
$F_5$ & [0.188, 0.675] & [0.511, 0.788] & [0.117, 0.519] & [0.064, 0.094] & [1.000, 1.000]\\
$F_3$ & [0.182, 0.639] & [0.556, 0.806] & [0.105, 0.532] & [0.073, 0.101] & [0.884, 0.958]\\
$H_1$ & [0.201, 0.616] & [0.564, 0.833] & [0.170, 0.560] & [0.069, 0.109] & [0.540, 0.822]\\
$H_2$ & [0.174, 0.611] & [0.607, 0.840] & [0.162, 0.573] & [0.065, 0.108] & [0.565, 0.838]\\
$H_3$ & [0.189, 0.607] & [0.548, 0.823] & [0.164, 0.566] & [0.066, 0.103] & [0.550, 0.832]\\
\bottomrule
\end{tabular}
\end{table}
\begin{table}[!htbp]
\centering\footnotesize
\caption{Paired differences with study-bootstrap intervals. Comparisons with $F_5$ use the declared 90\% level; other rows use 95\%. These are secondary endpoints, except the $H_3$ approximation gain reported in the main text.}
\begin{tabular}{lccc}
\toprule
Comparison & Sign accuracy & Captured gain & Within-task $r$\\
\midrule
$H_3-K$ & 0.038 [$-0.044$, 0.122] & 0.119 [0.014, 0.224] & 0.026 [$-0.120$, 0.213]\\
$H_2-K$ & 0.071 [$-0.012$, 0.155] & 0.118 [0.006, 0.224] & 0.020 [$-0.120$, 0.204]\\
$H_1-K$ & 0.046 [$-0.028$, 0.121] & 0.114 [0.014, 0.215] & 0.035 [$-0.104$, 0.217]\\
$H_3-F_5$ & 0.038 [$-0.011$, 0.092] & 0.050 [$-0.018$, 0.117] & $-0.066$ [$-0.129$, 0.017]\\
$H_2-F_5$ & 0.071 [0.022, 0.128] & 0.049 [$-0.009$, 0.115] & $-0.073$ [$-0.128$, 0.012]\\
$F_5-K$ & 0.000 [$-0.062$, 0.060] & 0.069 [$-0.037$, 0.168] & 0.092 [$-0.048$, 0.253]\\
$H_3-F_3$ & 0.012 [$-0.054$, 0.072] & 0.047 [$-0.062$, 0.148] & $-0.038$ [$-0.103$, 0.054]\\
\bottomrule
\end{tabular}
\end{table}

\begin{table}[!htbp]
\centering\small
\caption{D3 development results, not independent confirmation: 91 seen studies, 236 blocks, 2,195 effects. Source: \texttt{dev-20260923-022128.json}.}
\begin{tabular}{lrrrrrr}
\toprule
System & Within-task $r$ & Sign & Gain & MAE & Magnitude ratio & $r_F$\\
\midrule
Kernel & 0.552 & 0.895 & 0.420 & 0.067 & 0.61 & 0.863\\
$F_5$ & 0.607 & 0.939 & 0.529 & 0.079 & 1.15 & 1.000\\
$F_3$ & 0.592 & 0.937 & 0.504 & 0.084 & 1.21 & 0.979\\
$H_1$ & 0.533 & 0.927 & 0.409 & 0.080 & 1.14 & 0.897\\
$H_2$ & 0.551 & 0.927 & 0.446 & 0.081 & 1.15 & 0.911\\
$H_3$ & 0.561 & 0.933 & 0.496 & 0.079 & 1.13 & 0.917\\
\bottomrule
\end{tabular}
\end{table}
Development $H_3-K$ sign improvement is 0.038 \CI{0.008}{0.084}, gain 0.076 \CI{-0.017}{0.179}, and within-task $r$ difference 0.008 \CI{-0.087}{0.078}. The corresponding 90\% comparisons with $F_5$ are $-0.006$ \CI{-0.041}{0.007}, $-0.033$ \CI{-0.087}{0.028}, and $-0.047$ \CI{-0.101}{-0.009}. Kernel--flagship effect agreement is 0.863 on development and 0.608 on test, limiting extrapolation from developmental diagnostics. The test approximation primary is an improvement criterion, not the equivalence test proposed in an earlier planning document.

\section{Discrepancy model: full table}\label{app:d2}
\begin{table}[!htbp]
\centering\small
\caption{D2 parameter fits and bootstrap intervals. Selected seen studies use five model respondents per cell. The unselected fit uses 20 development studies with 50 respondents per cell. Source: \texttt{discrepancy\_model.json}.}
\begin{tabular}{lccc}
\toprule
Fit & $\beta$ & $\tau$ & $\sigma$\\
\midrule
Jev, selected seen & 0.925 [0.607, 1.173] & 0.037 [0.021, 0.046] & 0.080 [0.051, 0.128]\\
Astra, selected seen & 0.579 [0.317, 0.660] & 0.035 [0.020, 0.046] & 0.074 [0.047, 0.125]\\
Jev, unselected dev & 0.661 & 0.020 & 0.020\\
\bottomrule
\end{tabular}
\end{table}
\begin{table}[!htbp]
\centering\small
\caption{Study-weighted coverage and 90\% interval diagnostics. Test discrepancy intervals include parameter uncertainty; split-half results summarize the seen-study calibration. No-discrepancy intervals include human sampling uncertainty only. Interval score penalizes width and misses.}
\begin{tabular}{lrrrrr}
\toprule
Evaluation & 50\% & 80\% & 90\% & Width (90\%) & Score (90\%)\\
\midrule
Jev, split half & 0.581 & 0.808 & 0.882 & --- & ---\\
Astra, split half & 0.599 & 0.839 & 0.902 & --- & ---\\
Jev, test & 0.669 & 0.932 & 0.959 & 0.324 & 0.422\\
Astra, test & 0.625 & 0.859 & 0.937 & 0.295 & 0.394\\
Jev, no discrepancy & 0.140 & 0.288 & 0.362 & 0.064 & 0.746\\
Astra, no discrepancy & 0.204 & 0.334 & 0.397 & 0.064 & 0.754\\
Jev, unselected-dev fit on test & 0.349 & 0.582 & 0.665 & 0.116 & 0.472\\
\bottomrule
\end{tabular}
\end{table}
The likelihood fits arm effects rather than treating all dependent pairwise contrasts as independent. For interval $[l,u]$ at nominal $1-\alpha$, the interval score is $(u-l)+(2/\alpha)(l-y)\mathbf{1}_{y<l}+(2/\alpha)(y-u)\mathbf{1}_{y>u}$. Larger coverage alone can be bought with excessive width; the score makes that tradeoff visible. The selected-study fit and the unselected fit differ in both study selection and model sampling noise, so their comparison does not isolate either factor.

\section{Additional calibration and comparison tables}\label{app:extra}
\begin{table}[!htbp]
\centering\small
\caption{G-A's test decision-value curve. Entries reuse nested subsets of the same recorded kernel run and are not independent experiments. Costs are per simulated experiment at the recorded average of 4.3 conditions.}
\begin{tabular}{lrrrrrr}
\toprule
Respondents per cell & 5 & 10 & 20 & 50 & 100 & 200\\
\midrule
Sign accuracy & 0.655 & 0.674 & 0.632 & 0.645 & 0.661 & 0.671\\
Captured gain & 0.272 & 0.358 & 0.264 & 0.255 & 0.278 & 0.268\\
Cost (USD) & 0.0007 & 0.0015 & 0.0029 & 0.0074 & 0.0147 & 0.0295\\
\bottomrule
\end{tabular}
\end{table}

\begin{table}[!htbp]
\centering\small
\caption{Three-model comparison (A3/A3b), from \texttt{three\_models.json}. The intended common sample has 3,615 items; the historical Jev five-per-cell file lacks two. Upper-panel $r$ and magnitude ratio use the marginal scorer. Lower-panel $r$ uses comparable decision blocks; its paired differences therefore do not equal differences between all upper-panel entries.}
\label{app:models}
\begin{tabular}{lrrrrr}
\toprule
Model & $W_1$ & Condition $r$ & Magnitude ratio & Sign & Gain\\
\midrule
Jev p3 & 0.1360 & 0.323 & 0.89 & 0.645 & 0.272\\
GPT-5.6 Luna (low) & 0.1320 & 0.506 & 1.61 & 0.697 & 0.305\\
GPT-6 Astra (high) & 0.1013 & 0.489 & 1.58 & 0.671 & 0.336\\
\midrule
\multicolumn{6}{l}{Comparable-block $r$: Jev 0.347; Luna 0.495; Astra 0.489.}\\
\multicolumn{6}{l}{Astra--Jev $r$ difference: 0.142 [0.014, 0.289] (95\%).}\\
\multicolumn{6}{l}{Astra--Jev sign difference: 0.026 [$-0.031$, 0.083] (95\%).}\\
\multicolumn{6}{l}{Astra--Jev gain difference: 0.064 [$-0.048$, 0.168] (95\%).}\\
\bottomrule
\end{tabular}
\end{table}

\begin{table}[!htbp]
\centering\small
\caption{B2 memory replication. Magnitude is mean absolute inter-item Spearman correlation relative to humans; structure is correlation between the simulated and human correlation patterns. ``Distant/human'' compares distant-pair coherence after correction with people's. Source: B2 decision record.}
\begin{tabular}{lrrrrr}
\toprule
Study & Independent magnitude & Corrected & Independent structure & Corrected & Distant/human\\
\midrule
m52pd & 0.175 & 0.289 & 0.018 & 0.518 & 0.274\\
3xy9j & 0.548 & 0.645 & 0.005 & 0.263 & 0.611\\
gx6hp & 0.456 & 0.786 & 0.001 & 0.514 & 0.791\\
ef6my & 0.305 & 0.345 & 0.089 & 0.410 & 0.326\\
5mt6r & 0.338 & 0.523 & 0.132 & 0.711 & 0.471\\
ac9jm & 0.122 & 0.281 & $-0.067$ & 0.718 & 0.258\\
Median & 0.32 & 0.43 & 0.01 & 0.52 & 0.40\\
\bottomrule
\end{tabular}
\end{table}

\begin{figure}[!htbp]
\centering\includegraphics[width=\linewidth]{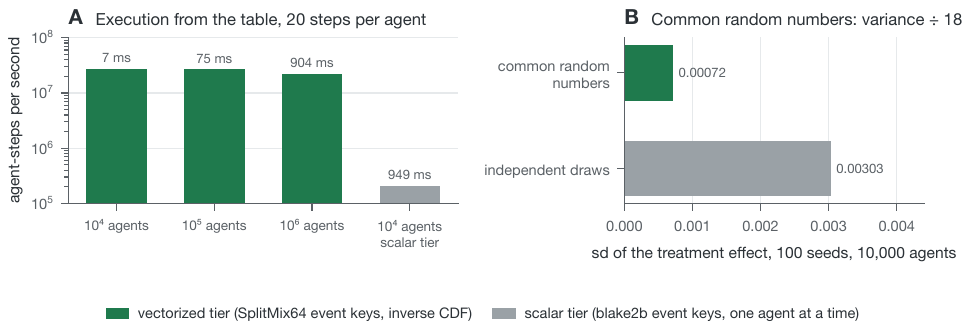}
\caption{Tabulated execution and variance reduction. A shared legend appears below the panels. (A) Agent-steps per second on a logarithmic scale for the vectorized engine at $10^4$, $10^5$, and $10^6$ agents and the scalar engine at $10^4$, with 20-step wall time above each bar. These timings exclude table construction and fresh inference. (B) Treatment-effect standard deviation across 100 seeds at $10^4$ agents: 0.00072 with common random numbers and 0.00303 with independent draws, an approximately 18-fold variance reduction. Bars summarize the recorded standard deviations, not seed-level distributions.}
\label{fig:scale}
\end{figure}

\section{Provenance and interpretation notes}
The included source packet separates frozen configuration, decision records, and aggregate JSON. Tables and Figures~\ref{fig:architecture}--\ref{fig:scale} use those records and the supplied vector figures. Per-effect human uncertainty in Table~\ref{tab:human-effects} and Figure~\ref{fig:epstein-detail} comes from the participant-level bootstrap resolved in \texttt{RESOLUTIONS.md}; no intervals are inferred from rounded means.

SocSci210's prepared 456,587 observations are response rows, and simulated sampling is capped per cell. We retain the data schema's distinction despite shorthand references to ``every respondent'' or ``people'' elsewhere. The three-model and D3 scorer populations differ, explaining their different condition-sensitivity correlations and magnitude ratios. The hybrid-minus-kernel sign-difference upper bound in D3 rounds to 0.122 in the JSON rather than 0.123 in the decision prose. D3's historical 90\% flagship comparisons also contain small rounding discrepancies; the appendix follows JSON rounding.

Under the p3 phrasing, each prediction issues one request with two option probes (``Nouls''); the ledger's \texttt{cache\_hits} and \texttt{cache\_misses} count probes, not predictions. Thus 11,680 probe hits represent 5,840 cached predictions (29.2\% of 20,000), and 28,320 probe misses represent 14,160 live predictions: 14,159 calls plus the one failed call completed on resume. The initial JSON records 19,999 predictions and that failure; the resumed run completes all 20,000. The D1 kernel cost was recovered from console output because a resumed ledger overwrote earlier entries; B2's approximately \$3 cost is estimated because its ledger was filtered out, and is not used as a measured cost claim.

External reference metadata were checked where accessible; APS is cited in its arXiv version and Ashokkumar et al.\ (2026) in its published version. Detailed editorial discrepancies, unresolved checks, title alternatives, categories, figure recipes, and a ten-line summary accompany the manuscript in \texttt{NOTES.md}.

\clearpage
\bibliographystyle{plainnat}
\bibliography{refs}
\end{document}